\documentclass[prl,twocolumn,showpacs,nofootinbib,superscriptaddress]{revtex4}
\usepackage{epsf}
\usepackage{epsfig}
\usepackage{amsmath}

\newcommand{\be}{\begin{eqnarray}}
\newcommand{\ee}{\end{eqnarray}}
\newcommand{\ba}{\begin{array}}
\newcommand{\ea}{\end{array}}

\begin{document}
\preprint{MKPH-T-08-04}

\title{Taming Deeply Virtual Compton Scattering}

\author{Maxim V. Polyakov}
\affiliation{Petersburg Nuclear Physics
Institute, Gatchina, St.\ Petersburg 188350, Russia}
\affiliation{Institut f\"ur Theoretische Physik II,
Ruhr--Universit\"at Bochum, D--44780 Bochum, Germany}

\author{Marc Vanderhaeghen}
\affiliation{Institut f\"ur Kernphysik, Johannes Gutenberg-Universit\"at,
D-55099 Mainz, Germany}
\affiliation{Physics Department, College of William and Mary,
Williamsburg, VA 23187, USA}

\date{\today}

\begin{abstract}
We study recent Deeply Virtual Compton scattering
(DVCS) data within a dual parameterization of the Generalized Parton
Distributions (GPDs). This parameterization
allows to quantify the maximum amount of information, that can be
extracted from DVCS data, in a ``quintessence'' function.
We present a ``zero step'' model for the latter
solely based on the forward quark density,
providing a parameter free prediction
for the imaginary part of the DVCS amplitude. It is shown that
the bulk effect of the $e p \to e p \gamma$ beam helicity
cross section difference can be understood within such a model.
\end{abstract}


\maketitle
\thispagestyle{empty}

The generalized parton distributions (GPDs) \cite{pioneers}
(see Refs.~\cite{GPV,Diehlrev,Belitskyrev,Boffi:2007yc} for recent reviews)
describe the response of the target hadron to the well-defined
QCD quark-gluon operator of the type~:
$ \bar \psi_\alpha(0)\ {\rm
P}e^{ig\int_0^z dx_\mu A^\mu}\ \psi_\beta (z)$,
which is defined on the light-cone, i.e. $z^2=0$.
The hard
exclusive processes provide us with a set of new fundamental
probes of the hadronic structure. The hadron image seen by such non-local
probes is encoded in the dependence of GPDs on its
variables.
The central quantities that contain invaluable information
on the nucleon structure are momentum transfer ($t$) dependent quark densities. These
densities can be obtained as the
 $\xi \to 0$ limit of the nucleon
GPDs $H(x,\xi,t)$ and $E(x,\xi,t)$~:
\be
q(x,t)=\lim_{\xi\to 0} H(x,\xi,t), \ \ \ e(x,t)=\lim_{\xi\to 0} E(x,\xi,t),
\label{limit}
\ee
where $x \pm \xi$ correspond to the quark longitudinal momentum fractions.
The first density $q(x,t)$ (at $t=0$ it reduces
to the usual quark distribution measured in DIS) is related to
the distribution of quarks and anti-quarks in the longitudinal
momentum and transverse plane of the nucleon, thus providing the 3D image
of quarks and anti-quarks in the nucleon \cite{Burkardt}. The new
function $e(x,t)$ is crucial for extraction of the angular momentum
carried by quarks in the nucleon \cite{Ji:1996}.

Although the quark densities $q(x,t)$ and $e(x,t)$
are defined as the simple limit of GPDs (\ref{limit}),
one can not perform this limit measuring observables
for hard exclusive processes. The reason is that
the leading order amplitude of hard exclusive reactions
(we restrict ourselves to DVCS and discuss only the GPD
$H(x,\xi,t)$, the discussion for $E(x,\xi,t)$ is
analogous) is expressed as~:
\be
A(\xi,t)=\int_{-1}^1 dx\,H(x,\xi,t)\, \left[\frac
1{\xi-x-i0} -\frac 1{\xi+x-i0}\right].
\label{elementaryAMP}
\ee
The amplitude is given by the convolution integral in
which the dependence of the GPDs on the variable $x$ is ``integrated out".
One cannot completely restore the
GPD $H(x,\xi,t)$ from~(\ref{elementaryAMP}),
hence one is not able to
perform a ``complete imaging" of the target hadron from the
knowledge of the amplitude and cannot perform the limit (\ref{limit}) to
obtain the key quark densities $q(x,t)$ and $e(x,t)$.

The aim of this Letter is to estimate the contribution of the quark
densities $q(x,t)$ to DVCS observables. This is to be considered as
a ``zero step" to extract the GPD information from the data, as it
allows us to quantify the deviations in terms of
genuine non-forward parts of the GPDs. For our analysis we employ the dual parameterization of GPDs suggested
in Ref.~\cite{MaxAndrei}.

The dual parameterization  is based on a
representation of parton distributions as an infinite series of
$t$-channel exchanges \cite{MVP98}, allowing to express
the GPD $H$ in terms of a set of functions
$Q_{2\nu}(x,t)$ ($\nu=0,1,2,\ldots$), for details see~\cite{MaxAndrei}.
We call the functions $Q_{2\nu}(x,t)$ {\it forward-like} because
at leading order (LO), the scale dependence of the functions
$Q_{2\nu}(x,t)$ is given
by the standard DGLAP evolution equation, so that these functions
behave as usual parton distributions under QCD evolution.
Furthermore,
the function $Q_0(x,t)$ is related to the forward $t$-dependent
quark densities $q(x,t)$ (which reduces at $t=0$ to the parton densities
$q(x)$ measured in DIS)~:
\be Q_0(x,t) = \left[ q + \bar q \right] (x,t)
-\frac x2  \int_x^1
\frac{dz}{z^2}\ \left[ q +\bar q \right](z,t) \, . \label{Q0}
\ee
The expansion of the GPD $H(x,\xi,t)$ around $\xi=0$
with $x$ fixed to the order $\xi^{2\nu}$ involves only a finite
number of functions $Q_{2\mu}(x,t)$ with $\mu\leq \nu$.
One can then express the amplitude (\ref{elementaryAMP}) in terms
of forward-like functions as~\cite{MaxAndrei,tomography,educing}:
\begin{widetext}
\begin{eqnarray}
{\rm Im\ } A(\xi,t)&=&
2 \, \int_{\frac{1-\sqrt{1-\xi^2}}{\xi}}^1 \frac{dx}{x} N(x,t)\ \Biggl[
\frac{1}{\sqrt{\frac{2 x}{\xi}-x^2-1}}
\Biggr]\, ,
\label{IM}
\\
\nonumber {\rm Re\ } A(\xi,t)&=&
2 \, \int_0^{\frac{1-\sqrt{1-\xi^2}}{\xi}} \frac{dx}{x} N(x,t)\ \Biggl[
\frac{1}{\sqrt{1-\frac{2 x}{\xi}+x^2}} + \frac{1}{\sqrt{1+\frac{2
x}{\xi}+x^2}}-\frac{2}{\sqrt{1+x^2}}
\Biggr]  \\
&+& 2 \, \int^1_{\frac{1-\sqrt{1-\xi^2}}{\xi}} \frac{dx}{x} N(x,t)\
\Biggl[ \frac{1}{\sqrt{1+\frac{2
x}{\xi}+x^2}}-\frac{2}{\sqrt{1+x^2}} \Biggr]+ 4 D(t) \, .
\label{RE}
\end{eqnarray}
\end{widetext}
Here $D(t)$ is the $D$-form factor (FF)~:
\be D(t)=\sum_{n=1}^\infty
d_n(t)=\frac 12 \int_{-1}^1 dz\ \frac{D(z,t)}{1-z}\, , \ee where
$D(z,t)$ is the $D$-term \cite{PW99}. One can check
\cite{tomography} that the amplitude given by Eqs.~(\ref{IM},\ref{RE})
automatically satisfies a dispersion relation with the
subtraction constant given by the $D$-FF, as it should be
on general grounds \cite{Teryaev,DiehlIvanov}. The function
$N(x,t)$ in Eqs.~(\ref{IM},\ref{RE}) is defined as~:
\be N(x,t)=\sum_{\nu=0}^\infty x^{2\nu}\ Q_{2\nu}(x,t)\, .
\label{QF} \ee
The information contained in the LO amplitude is
in one-to-one correspondence with
the knowledge of the function $N(x,t)$ and the $D$-FF
$D(t)$, because Eq.~(\ref{IM}) can be inverted
\cite{tomography}, i.e. the function $N$ can be expressed
{\it unambiguously} in terms of the amplitude. This inversion
corresponds to the Abel transform tomography \cite{Abel}, and
the corresponding inversion equation has the form \cite{tomography}~:
\be
N(x,t)&=&\frac{1}{\pi}\ \frac{x(1-x^2)}{~~~(1+x^2)^{3/2}}
\int_\frac{2 x}{1+x^2}^1\frac{d\xi}{\xi^{3/2}}\
\frac{1}{\sqrt{\xi-\frac{2 x}{1+x^2}}}
\nonumber \\
&& \hspace{.7cm}\times \left\{ \frac 12{\rm Im\ }
A(\xi,t)-\xi \frac{d}{d\xi}{\rm Im\ } A(\xi,t) \right\}.
\label{nina} \ee
%
\indent
This equation implies that
the function $N(x,t)$ contains the maximal
information about GPDs that is possible to obtain from the
observables. Therefore this function can be called the {\it GPD-quintessence
function} \footnote{See Refs.~\cite{tomography,educing} for the
detailed discussion of properties and physics interpretation of
the GPD-quintessence function.}.
Another important feature of the
expressions (\ref{IM},\ref{RE}) for the amplitude is that one can easily
single out  the contributions to the amplitude coming from the
forward parton densities. Indeed, the first term in the sum
(\ref{QF}) is given by the function $Q_0$ which is related to
the ($t$-dependent) parton densities by Eq.~(\ref{Q0}).
The big advantage of the dual parameterization is that
one can clearly separate the contribution of the ($t$-dependent)
parton densities from genuine non-forward effects encoded in
the functions $Q_2, Q_4, \dots$.
The authors of Ref.~\cite{Kumericki:2007sa} developed an approach that
also allows
 to separate the contribution of forward quark
densities to observables. Calculations of DVCS observables
in the dual parameterization were presented in~\cite{Guzey},
however in this paper a ``predefined" model for
forward-like functions $Q_{2 \nu}$
has been used and the ``zero step" separation was not discussed.

In the following we study the separation between effects of
forward quark densities and genuine non-forward contributions.
In order
to make this separation (``zero step") as clean as possible we choose to
analyze recent  JLab/Hall~A data \cite{MunozCamacho:2006hx}
on the beam helicity (in)dependent cross sections
in the $e+p\to e'+p+\gamma$ process, as well as
recent beam spin asymmetry data
measured by the CLAS collaboration \cite{girod:2007jq}.
We make such choice because the beam helicity dependent cross sections
and beam asymmetries are directly proportional
to the imaginary part of the DVCS amplitude.
This, in principle, gives the possibility to
apply directly the Abel tomography formula (\ref{nina}),
and requires a measurement at several values of $x_B$.
Furthermore, we choose the data with $Q^2>2$~GeV$^2$ and $-t<0.3$~GeV$^2$.
In this kinematical region one can safely neglect contributions
of the nucleon GPDs $E$, $\widetilde H$ and $\widetilde E$ relative
to the contribution of $H$, also one keeps the contribution of higher twists
negligible.
This allows us to make direct conclusions about contribution of the
quark densities $q(x,t)$. The region of small $t$ makes our results
insensitive to the modelling of the $t$-dependence of the quark densities.

In order to make this ``zero step" comparison of
the data with the contribution of forward quark densities we have
to make assumptions about the $t$-dependence of the quark densities
$q(x,t)$. As the model for this
dependence we use the Regge motivated Ansatz~:
$q(x, t) \,=\, q(x) \cdot x^{- \alpha^\prime \, t}$,
where $q(x)$ the forward quark distribution, and the slope
parameter $\alpha^\prime = 1.105$~GeV$^{-2}$ was fixed from the
form factor sum rule in Ref.~\cite{ansatz}.
\begin{figure}
\includegraphics[width =9.cm]{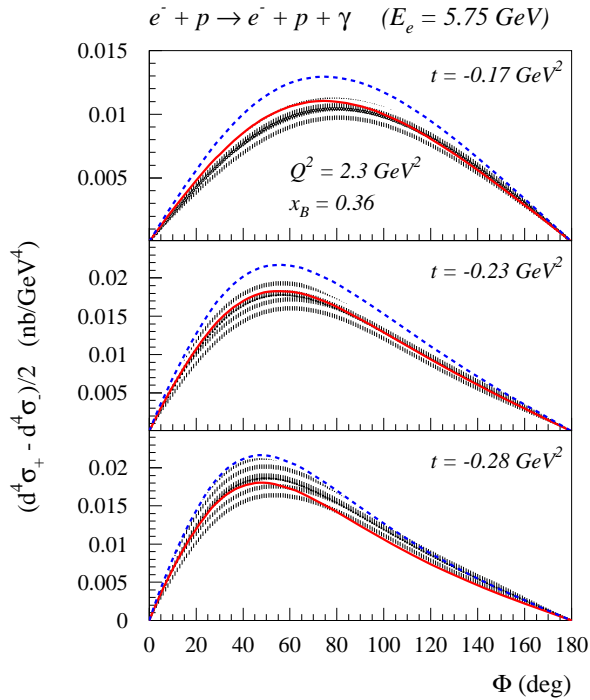}
\caption{Azimuthal angular dependence of the $e^- p \to e^- p
\gamma$ polarized cross section difference $(d^4 \sigma_+ - d^4
\sigma_-)/2$ for different beam helicities. The black bands are
JLab/Hall A data~\cite{MunozCamacho:2006hx}.
Dashed (blue) curves~: double distribution parameterization with profile
parameter $b = 1$;
solid (red) curves~: dual parameterization including
only the forward function $Q_0$.}
\label{fig:halla}
\end{figure}

\begin{figure}
\includegraphics[width =9.cm]{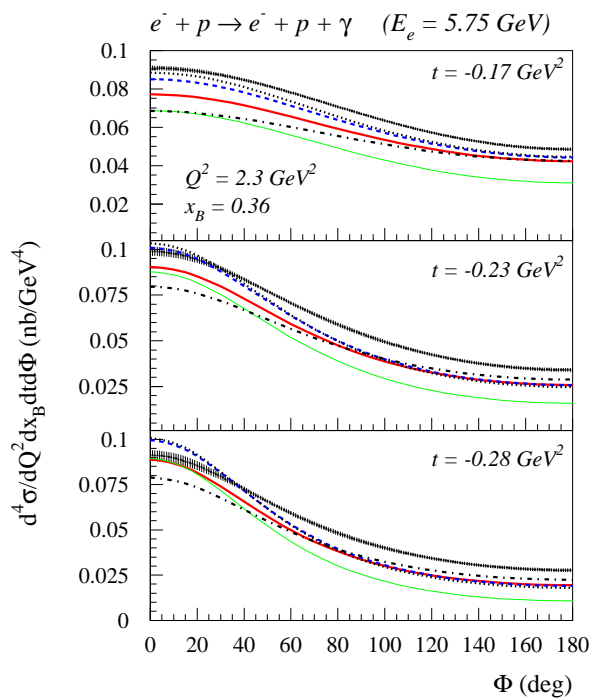}
\caption{Azimuthal angular dependence of the $e^- p \to e^- p
\gamma$ unpolarized cross section. The black bands are JLab/Hall A
data~\cite{MunozCamacho:2006hx}.
The dashed blue curves show the double distribution result with profile
parameter $b = 1$.
The results in the dual parameterization including
only $Q_0$ are shown for
$D(0) = 0$ (thick solid red curves),
$D(0) = -4/3$ (dashed-dotted curves),
and $D(0) = +4/3$ (dotted curves).
The lower thin solid green curves are the Bethe-Heitler cross section.}
\label{fig:hallaunpol}
\end{figure}

In Fig.~\ref{fig:halla}, we compare our predictions for the
polarized cross section difference for different beam helicities 
for which the first data have been
published by the JLab/Hall A Coll.~\cite{MunozCamacho:2006hx}.
This observable is directly proportional to the
imaginary part of the DVCS amplitude.
The data are compared with predictions computed in the dual parameterization
and in a double distribution model.
For both models we use the same
Regge-type $t$-dependence for forward $t$-dependent quark densities. 
For the double distribution we use the
profile function with parameter $b = 1$ as
in estimates performed in Refs.~\cite{Vanderhaeghen:1999xj,GPV}.
It is seen that the double distribution model yields polarized cross sections
that have a tendency to be somewhat larger than the data.
When using only the forward function $Q_0$ in the dual parameterization,
it is seen from Eq.~(\ref{Q0}) that the
imaginary part of the DVCS amplitude yields a parameter free
prediction at $t = 0$ \footnote{We checked that in the $t$ range
shown, the dependence on $\alpha^\prime$ is much
smaller than the difference between the curves.}.
It is seen that this parameter free prediction yields an amazingly consistent
description of the polarized cross sections.

The JLab/Hall A Coll. also published results for unpolarized
cross sections that are shown in Fig.~\ref{fig:hallaunpol}. When
comparing the unpolarized cross sections with the model
independent Bethe-Heitler result, one sees that the latter dominates
the cross section at $\Phi = 0$, where it makes up for 
about 85 \% of the cross section, at $-t = 0.23$~GeV$^2$. 
However, at $\Phi = 180$~deg it is more than a factor 2 below the
data.
Although both double distribution and dual parameterization models
can explain part of the difference with the data,
no single model is able to simultaneously describe the cross section
at $\Phi = 0$ and $\Phi = 180$~deg,
in line with the finding of~\cite{guidal2008}.
In particular, within the dual parameterization in twist-2 approximation,
adjusting the one subtraction constant $D(t)$ does not allow to 
describe this $\Phi$ dependence,
as is also shown on Fig.~\ref{fig:hallaunpol}.
It was also checked that when adding an estimate for
the non-forward function $Q_2$ no agreement can be found either over the
whole $\Phi$ range.
Although the region around $\Phi = 180$~deg yields zero for
the beam helicity asymmetry, it is very
worthwhile to cross-check this puzzle further.

\begin{figure}
\includegraphics[width =8.cm]{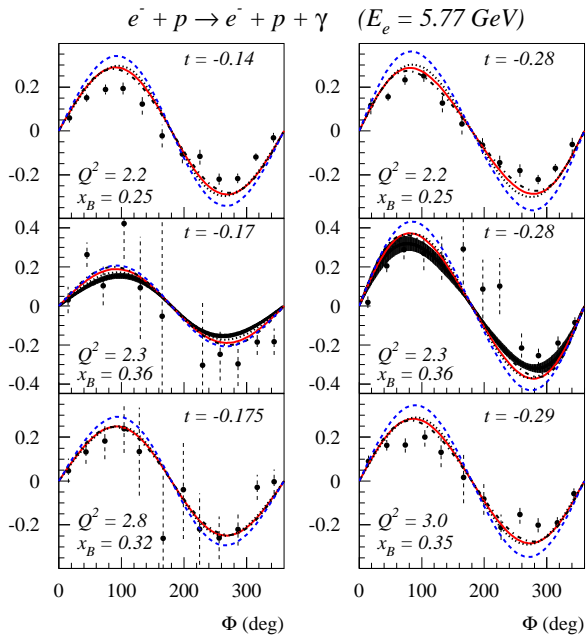}
\caption{Azimuthal angular dependence of the $e^- p \to e^- p
\gamma$ beam helicity asymmetry $(d^4 \sigma_+ - d^4
\sigma_-)/(d^4 \sigma_+ + d^4 \sigma_-)$ for different kinematics.
Black bands in the two middle
panels are data points from JLab/Hall
A~\cite{MunozCamacho:2006hx}.
Solid circles are data points from
JLab/CLAS~\cite{girod:2007jq}.
Curve conventions as in Fig.~\ref{fig:hallaunpol}
}
\label{fig:clas6}
\end{figure}

In Fig.~\ref{fig:clas6}, we show recent exclusive measurements of
 $e^- p \to e^- p \gamma$ beam
helicity asymmetries from JLab/CLAS~\cite{girod:2007jq} and
JLab/HallA~\cite{MunozCamacho:2006hx}.  We note that for
two middle  bins in Fig.~\ref{fig:clas6}, 
where both experiments have overlapping kinematics, 
the data from both experiments are consistent with each other. 
When comparing both parameterizations with the data in Fig.~\ref{fig:clas6},
we note that the double distribution model (dashed curves) yields
asymmetries that lie systematically above the data. The dual
parameterization model based on the forward function $Q_0$ (solid
curves) yields a good first (``zero step") description of the data,
given that {\it no} parameter was adjusted here. There is a slight tendency
 for the asymmetries to be overestimated within the dual parameterization
which is merely a reflection of the underestimate of the
unpolarized cross sections as seen in Fig.~\ref{fig:hallaunpol}.

The success of our ``zero step" exercise shows
that the contribution of the forward quark densities $q(x,t)$
at small $t$
to the $e^- p \to e^- p
\gamma$ polarized cross section difference yields the bulk effect. Deviations of this prediction from the data
 can be fitted by introducing forward-like functions $Q_2,Q_4, \ldots$
 which describe the
 genuine non-forward effects in GPDs.
The dominance of the forward quark densities in 
the imaginary part of the DVCS amplitude is an important observation. 
It implies that precise
measurements of DVCS observables in a wider kinematical region
would allow us to extract the $t$-dependent
quark densities $q(x,t)$ and $e(x,t)$. The latter can be
accessed by measurements on the neutron.

In summary, we studied recently obtained high precision DVCS data
in the valence region within a dual parameterization of the GPDs.
This parameterization allows to extract a quintessence function
which contains the maximal amount of information that can
be extracted from DVCS observables.
We established a ``zero step'' model for this function solely based on the
forward quark density, which provides a parameter free prediction
for the imaginary part of the DVCS amplitude. It was shown that the
bulk effect of the $e p \to e p \gamma$ beam helicity
cross section difference can be understood within such a model.
By systematically studying deviations
between the data and such model will allow to reveal the
non-forward effects to this quintessence function.

\begin{acknowledgments}

The work of M.V. is supported in part by the
U.S.\ Department of Energy grant no.\ DE-FG02-04ER41302.
The work of M.V.P. is supported in part by the AvH
Foundation, BMBF and DFG.
The authors thank M. Guidal and D.~M\"uller for useful discussions
and C. Munoz Camacho and F.X. Girod for sending the data.

\end{acknowledgments}

\end{document}